# Continuously tunable anomalous Hall crystals in rhombohedral heptalayer graphene

Hanxiao Xiang[1,2,*], Jing Ding[1,2,*,†], Jiannan Hua[1,2,*], Naitian Liu[1,2], Wenqiang Zhou[1,2], Qianmei Chen[3], Kenji Watanabe[4], Takashi Taniguchi[5], Na Xin[3], Wei Zhu[1,2,†], Shuigang Xu[1,2,†]

[1] *Key Laboratory for Quantum Materials of Zhejiang Province, Department of Physics, School of Science, Westlake University, Hangzhou 310030, China*

[2] *Institute of Natural Sciences, Westlake Institute for Advanced Study, Hangzhou 310024, China*

[3] *Department of Chemistry, Zhejiang University, Hangzhou 310058, China*

[4] *Research Center for Electronic and Optical Materials, National Institute for Materials Science, Tsukuba 305-0044, Japan*

[5] *Research Center for Materials Nanoarchitectonics, National Institute for Materials Science, Tsukuba 305-0044, Japan*

[*]These authors contributed equally to this work.
[†]Corresponding authors: dingjing@westlake.edu.cn (J.D.); zhuwei@westlake.edu.cn (W.Z.); xushuigang@westlake.edu.cn (S.X.)

**Abstract:**
The interplay of electronic interactions and nontrivial topology can give rise to a wealth of exotic quantum states. A notable example is the formation of Wigner crystals driven by strong electron-electron interactions[1-8]. When these electronic crystals emerge in a parent band carrying a large Berry curvature, they can exhibit topologically nontrivial properties as anomalous Hall crystals, spontaneously breaking both continuous translational symmetry and time-reversal symmetry[9-13]. Here, we report the experimental observation of tunable anomalous Hall crystals in rhombohedral heptalayer graphene moiré superlattices. At filling factors near one electron per moiré unit cell ($\nu = 1$), we identify a series of incommensurate Chern insulators with a Chern number of $C = 1$. Furthermore, we observe spontaneous time-reversal symmetry breaking spanning the entire filling range from $\nu = 1$ to $\nu = 2$, manifesting as anomalous Hall effects with pronounced magnetic hysteresis. Notably, anomalous Hall crystals with a high Chern number $C = 3$ are observed over generic fillings ranging from $\nu = 1.5$ to $\nu = 2$. These anomalous Hall crystals are incommensurate with the moiré superlattice and exhibit dispersive fan diagrams consistent with the Streda formula, with their positions continuously tunable through displacement fields. Remarkably, these partially filled Chern insulators display Chern numbers distinct from their parent bands. Our findings demonstrate the rich variety of electronic crystalline states in rhombohedral graphene moiré superlattices, offering valuable insights into the strongly correlated topological phases.

**Main text:**

Atoms can arrange themselves into highly ordered microscopic structures through chemical bonds, forming various types of crystals that exhibit abundant physical properties. Analogously, electrons can crystallize through strong correlations by spontaneously breaking continuous translational symmetry, leading to the emergence of exotic quantum states. A prominent example of such electron crystallization is the Wigner crystal, which arises in strongly interacting electron systems with trivial topology and is manifested as insulating states with zero Hall conductivity[1-8]. In topologically nontrivial systems, interaction-driven topological analogs of Wigner crystals, known as Hall crystals, can emerge. The concept of Hall crystals was originally proposed within the conventional quantum Hall regime under a high magnetic field $B$[14]. Theoretically, Hall crystals can also form in the absence of a magnetic field by spontaneously breaking both continuous translational symmetry and time-reversal symmetry, resulting in what are known as anomalous Hall crystals (AHC)[9-13]. AHC can exhibit quantized anomalous Hall conductance at unexpected filling factors, with the Chern number of these states potentially differing from that of the parent band[12].

Recent experimental observations of integer and fractional quantum anomalous Hall (QAH) effects in rhombohedral multilayer graphene moiré superlattices have elevated these systems into fascinating platforms for engineering the interplay of correlations and topology[15-21]. Surprisingly, these states appear at strong displacement fields, where electrons are pushed away from the moiré interface such that an isolated moiré band is absent in the single-particle picture (see our calculation in Fig. 1i). The formation of AHC has been proposed as a candidate to explain these puzzling results, but direct experimental evidence has remained elusive[9-13]. Furthermore, earlier studies show that QAH states predominately emerge when they are commensurate with the moiré superlattices, occurring only at specific filling factors $\nu$ (defined as the number of electrons per moiré unit cell)[15,16,19,20]. Meanwhile, although other moiré systems have reported topological electronic states in non-integer fillings of the moiré band, such as topological charge density waves, electronic crystallization still appears to be commensurate with the moiré superlattices[22,23]. In contrast, the theoretical formation of AHC arises purely from electron interactions, with the moiré potential providing only a pinning effect. Hence, AHC can persist over a continuous range of filling factors by adjusting lattice constants and can survive even without the moiré potential[10].

Here, we report the observation of AHC in rhombohedral heptalayer graphene and hexagonal boron nitride (h-BN) moiré superlattices near $\nu = 1$ and extending through the entire region from $\nu = 1$ to $\nu = 2$. These results are revealed through the Streda formula dispersion and anomalous Hall effect measurements. Near $\nu = 1$, we observe quantized anomalous Hall resistance $R_{xy} = h/e^2$, accompanied by a series of dispersive fan diagrams emanating from different $\nu$ values, but all sharing the same Chern number $C = 1$. Between $\nu = 1$ and $\nu = 2$, we observe spontaneous time-reversal symmetry breaking across the entire range, evidenced by anomalous Hall effects with pronounced hysteresis. The topological states in this range can be classified into two types: $C = 1$ and $C = 3$. The corresponding $\nu$ of these emergent Chern states can be continuously tuned by the displacement field, as revealed through the Streda formula and maximum anomalous Hall resistance, indicating the tunable unit cell size. While very recent data on pentalayer and tetralayer graphene offer a hint towards the existence of topological electronic crystals[18,20], the heptalayer system studied here provides strong evidence supporting the existence of AHC, illustrated by a series of the Streda slopes near $\nu = 1$ and atypical topological crystals in the second correlated band.

**Quantum anomalous Hall insulator**

The surface flat band and large parent Berry curvature of rhombohedral graphene provide a fertile platform for exploring strong correlations and topological phenomena[21,24-30]. The moiré superlattice formed at graphene/h-BN interfaces further flattens the low-energy bands, with the layer number serving as an additional tunable parameter. We fabricated rhombohedral heptalayer graphene moiré superlattices by crystallographically aligning graphene with h-BN. Figure 1a illustrates the device structure. Dual gate structures with a bottom graphite gate and a top metallic gate enable independent tuning of carrier density $n$ and perpendicular displacement field $D$. We have measured two devices from separate heterostructures, exhibiting similar phenomena. In the main text, we mainly present data from device D1, with device D2 results shown in Extended Data Figs. 10 and 11.

The twist angle between graphene and h-BN for device D1 was calculated as 0.53° (see Methods). Figure 1b,c present the phase diagram of device D1, showing Hall resistance $R_{xy}$ and longitudinal resistance $R_{xx}$ as functions of $\nu$ and $D$ at $T = 50$ mK. We focus on regions where electrons are polarized into the moiréless interface, producing diverse topological states. A complete map of $R_{xx}$ is available in Extended Data Fig. 2. To mitigate cross-mixing between $R_{xx}$ and $R_{xy}$ and to suppress the zero-field magnetic fluctuations, $R_{xx}$ and $R_{xy}$ have been symmetrized and anti-symmetrized at $B = \pm 50$ mT, respectively. Near $\nu = 1$, a large $R_{xy}$ appears accompanied by local $R_{xx}$ minima, indicating a large Hall angle. The emergence of large anomalous Hall signals indicates the system exhibits spontaneous time-reversal symmetry breaking, further confirmed by magnetic hysteresis loops upon sweeping $B$. Figure 1f shows $R_{xy}$ and $R_{xx}$ as a function of $B$ swept forward and backward at fixed $\nu = 1$ and $D = 0.640$ V nm$^{-1}$. The remarkable hysteresis loop confirms the presence of ferromagnetic order with a coercive field around 12 mT at base temperature. The temperature-dependent hysteresis yields a Cuire temperature of 3.0 K (see Extended Data Fig. 9). Both are in line with previous reports on other rhombohedral graphene moiré systems[15,16].

Notably, in Fig. 1f, at base temperature and zero field, $R_{xy}$ reaches 25.4 kΩ, nearly quantized at $h/Ce^2$ with $C = 1$. Meanwhile, $R_{xx}$ is smaller than 1 kΩ. Figure 1g shows $R_{xx}$ and $R_{xy}$ as a function of $\nu$ at fixed $D = 0.644$ V nm$^{-1}$. A plateau feature can be observed near $\nu = 1$. The quantized plateau of $R_{xy}$ can also be seen by plotting $R_{xy}$ and $R_{xx}$ as a function of $D$ at fixed $\nu = 1$, as shown in Fig. 1h. All these features confirm the observation of QAH effect at $\nu = 1$. Besides the quantized value, the Chern number of QAH insulator can also be determined through the Streda formula $\frac{\partial n}{\partial B} = Ce/h$[31,32]. Figure 1d,e display Landau fan diagrams mapping $R_{xy}$ and $R_{xx}$ against $\nu$ and $B$ near $\nu = 1$. Pronounced dispersion with $B$ in the minima of $R_{xx}$ and plateau of $R_{xy}$ down to $B = 0$ confirms the topologically nontrivial states. The slopes of the dispersion marked by dashed lines yield a Chern number of $C = 1$ according to the Streda formula, consistent with the quantized value. Our experimental observations of topological gap at $\nu = 1$ agree with interacting band structure using self-consistent Hartree-Fock calculations as shown in Fig. 1j, confirming the important role of electron interactions in isolating the Chern band.

**Incommensurate Chern insulators**

Increasing $D$ from 0.640 V nm$^{-1}$ to >0.650 V nm$^{-1}$, we find multiple stripe features in the fine maps of $R_{xx}$ and $R_{xy}$ near $\nu = 1$, as shown in Fig. 2a,b. Line cuts along the dashed lines marked in Fig. 2a,b are shown in Fig. 2c. Intriguingly, $R_{xx}$ oscillates as a function of $\nu$. Meanwhile, $R_{xy}$ repeatedly exhibits local plateau-like features. These features were absent in previous tetralayer, pentalayer and hexalayer graphene systems[15,16,18,19]. Figure 2d,e shows the scan of $B$ forward and backward, acquired at the positions corresponding to the dips marked in Fig. 2b. These states exhibit remarkable hysteresis loops with $R_{xy}$ close to the quantized value of $h/e^2$. To further confirm their topologically nontrivial characteristics, we measured the Landau fan diagrams of $R_{xx}$ and $R_{xy}$ near $\nu = 1$ with fixed $D$ =0.675 V nm$^{-1}$, as shown in Fig. 2f,g. Remarkably, all the states disperse with $B$ as depicted in the corresponding Wannier diagram as shown in Fig. 2h. They share the same Streda slopes with $C = 1$, which are consistent with the value based on quantized $R_{xy}$. These subtle features can be smeared at $T = 2$ K and the main emanating positions are pinned to $\nu = 1$ (see Extended Data Fig. 5). This suggests incommensurate electronic states are easily melting at elevated temperatures[6], while moiré potential facilitates pining them at commensurate fillings[3].

Our observations of a series of Chern insulators near $\nu = 1$ indicate topological states can occur at incommensurate moiré fillings. This contradicts the mechanism of Chern insulators driven by interaction-induced valley-polarized Chern bands[33,34], which accounts for most Chern insulator systems such as twisted graphene and twisted MoTe$_2$[35-40]. In contrast, our results consist with the AHC mechanism recently proposed to explain the lack of an isolated flat band in rhombohedral graphene moiré systems[9-13]. In this picture, it is purely the electronic interactions that drive the spontaneous continuous translational and time-reversal symmetry breakings simultaneously through electronic crystallization, while the moiré superlattice is not an essential factor. Therefore, the Chern states can be incommensurate with the moiré superlattices through the redistribution of electron density during the crystallization. Our data, together with the very recent observations of extended QAH states at $\nu < 1$ in rhombohedral penta- and tetra-layer graphene systems[18], support the AHC mechanism, which can explain the puzzling topological gap at $\nu = 1$. Furthermore, we observe stronger evidence for the existence of AHC between $\nu = 1$ and $\nu = 2$, which has not been reported previously and will be discussed in the following sections.

**Anomalous Hall effects at second correlated band**

Besides the unusual behavior of the topological state near $\nu = 1$, we observe previously unexplored topological states between $\nu = 1$ and $\nu = 2$. Figure 3a,b display the fine color map of the symmetrized $R_{xx}$ and anti-symmetrized $R_{xy}$ acquired at $B = \pm 50$ mT in this region. Clearly, non-vanishing $R_{xy}$ and corresponding dips in $R_{xx}$ can be identified at a specific $D$ for each $\nu$. The anomalous Hall effect is further confirmed from the magnetic hysteresis loops as shown in Fig. 3f by sweeping $B$ forward and backward at various fixed $\nu$ and $D$. Particularly, near $D = 0.684$ V nm$^{-1}$, $R_{xy}$ is close to a quantized value of $h/3e^2$ in the range of $1.57 \leq \nu \leq 1.68$, as shown both in Fig. 3c and Fig. 3d. The observed high-Chern number QAH insulator with $C = 3$ at $\nu = 1.63$ (Fig. 3c), together with QAH insulator with $C = 1$ at $\nu = 1$ (Fig. 1f), suggests the abundant topological states in our system. The nonlinear $R_{xy}$ gradually smears with increasing temperature and eventually normal Hall effect is observed at $T > 3.0$ K (see Extended Data Fig. 9).

Figure 3e summaries the anomalous residual Hall resistance $\Delta R_{xy}$ as functions of $\nu$ and $B$ measured along the dashed lines marked in Fig. 3a,b. Intriguingly, the observed ferromagnetism spans the entire region from $\nu = 1$ to $\nu = 2$, although the amplitude of anomalous Hall signals fluctuates with $\nu$. Notably, the anomalous Hall effect is observed at generic $\nu$ incommensurate with the moiré superlattice, indicating the different origin from the common spontaneous valley polarization mechanism. The maximum anomalous Hall signals at each $\nu$ are optimized by tuning $D$. At a fixed $D$, there are three extreme $R_{xy}$. Both the non-vanishing $R_{xy}$ at low field and anomalous Hall effects with magnetic hysteresis indicate spontaneous time-reversal symmetry breaking occurs across the entire second correlated band. We found that such anomalous Hall effects at the second correlated band are reproducible in device D2 with a slightly different twist angle (see Extended Data Fig. 10).

**Continuously tunable Chern insulators at generic fillings**

The observed ferromagnetism in the region between $\nu = 1$ and $\nu = 2$ has a topological origin, which can be further revealed according to the Streda formula in the Landau fan diagrams. Figure 4 shows the evolution of the fan diagrams with various $D$. At fixed $D = 0.675$ V nm$^{-1}$, besides $\nu = 1$, there are additional three $\nu$ emanating dispersive features as shown in Fig. 4c, which are consistent with those in Fig. 4a. The corresponding $R_{xy}$ exhibit non-vanishing value at $B = 0$ and sign reversal across $B = 0$, associated with dips in $R_{xx}$ (see Extended Data Fig. 6). All these features are consistent with topologically nontrivial characteristics. The corresponding Chern states are annotated with the intercept filling and the Streda slope $(\nu, C)$. Notably, the $\nu$ from which the Streda slopes emerge are incommensurate with the moiré fillings. Even more surprisingly, $\nu$ can be continuously tuned by $D$ as demonstrated in Fig. 4c-f. Figure 4b plots the detailed evolution of $\nu$ with $D$ for the four groups of Chern states, in which one displays $C = 3$ and the remaining three exhibit $C = 1$. Such continuous tunability of the Chern insulators has not been observed in other moiré systems.

Intriguingly, the Landau fan at $\nu = 2$ indicates it is a topological trivial state ($C = 0$), manifesting as very large resistive $R_{xx}$, nondispersive $R_{xy}$ with no sign changes across $B = 0$, and absence of anomalous Hall effect (Extended Data Figs. 6 and 8). In Extended Data Fig. 8, we demonstrate it is likely to be an incommensurate Wigner crystal. This means that the topologically nontrivial Chern states at non-integral $\nu$ of the second correlated band have different topology from that of their parent band at $\nu = 2$.

**Discussion**

Our observations of incommensurate Chern insulators near $\nu = 1$ and in the second correlated band enrich the topological phases identified in rhombohedral multilayer graphene moiré systems. These findings contribute to unraveling the mysterious underlying mechanisms of the diverse Chern states emerging in these systems. Previously, to understand the contradiction between the experimental observations of QAH insulator at $\nu = 1$ and the absence of a topological gap in single particle theory, several models have been proposed[9-13,41,42]. Among these, the AHC mechanism stands out. Unlike other Chern insulators formed by correlation-induced spontaneous valley polarization, the AHC arises from electronic crystallization involving a spontaneous redistribution of electron density that opens a topological gap. In the AHC picture, it is mainly strong correlations, rather than moiré potentials, that fold the original flat bands into a smaller Brillouin zone and play an essential role in determining the Chern numbers of the folded bands[13]. Our experimental data support the AHC picture through the following aspects.

Firstly, the incommensurate topologically nontrival states observed near $\nu = 1$ and between $\nu = 1$ and $\nu = 2$ are distinct from previously reported generalized Wigner crystal and commensurate topological electronic crystals. The latter emerge only at specific $\nu$ commensurate with the moiré superlattices[15,16,19,20,22,23,43,44]. Additionally, we identify incommensurate Wigner crystals for $\nu < 0.5$ and near $\nu = 2$ (Extended Data Figs. 7 and 8). These abundant incommensurate states align with the AHC picture, in which electronic crystallization continuously adjusts the unit cell size to maintain one electron per unit cell for a given $n$ (or $\nu$). Our experimental data identify the $D$ as the primary control parameter governing these adjustments. The subtle variation of $D$ can modify the band structure and Berry curvature distribution. This tunability enables precise targeting of gap-instability hot spots in momentum space, such as Fermi surface nesting points, favoring a series of AHC at continuous fillings.

Secondly, the incommensurate Chern states with $C = 1$ near $\nu = 1$ could be attributed to the formation of Wigner crystals of holes (for $0.9 < \nu < 1$) or electrons (for $1 < \nu < 1.1$) superimposed on a topological $\nu = 1$ state, since we can simultaneously observe a series of Chern states at a fixed $D^{45}$. However, this mechanism cannot account for the Chern states observed at $\nu > 1.5$. States in this regime exhibit a different Chern number ($C = 3$) compared to their parent band ($C = 0$ at $\nu = 2$). The AHC model accommodates these observations, as interactions can modify the distribution of the Berry curvature, giving rise to Chern states that differ from the parent band[13].

Thirdly, notably, the incommensurate QAH states, as shown in Fig. 2, are more remarkable when $|D|$ is larger than that in Fig. 1. This suggests that weakening the influence of the moiré potential promotes the formation of incommensurate Chern insulators. Weaker moiré potentials appear to de-pin Chern states from commensurate fillings (e.g., $\nu = 1$ in our device). This behavior is consistent with the AHC picture. Our results indicate that in rhombohedral graphene moiré superlattices, the primary role of the moiré potential is to flatten the bands and enhance correlations rather than isolate topological bands. A strong moiré potential competes with the AHC and eventually destabilizes it, as evidenced by the absence of any topological states on the strong moiré interface[21].

**Outlook**
Our findings highlight a rich landscape of topologically correlated states arising from the interplay of electronic topology and spontaneous translational symmetry breaking. The observed AHC establish an entirely novel mechanism for the formation of Chern bands. Importantly, the displacement field offers a unique way to *in-situ* optimize electronic crystallization. The discovery of Chern insulators in the second correlated band, mimicking higher Landau levels, introduces a promising platform for exploring exotic QAH states, such as even-denominator fractional QAH states harboring non-Abelian anyons[46,47]. Additionally, the AHC with a high Chern number offer a basis for exploring topological phases beyond the traditional Landau level paradigm[48,49]. The distribution of Berry curvature varies with the layer number and twist angle in rhombohedral graphene moiré superlattices, potentially yielding diverse ground states. This variability offers many opportunities for exploring more exotic states in future studies.

**Methods**

**Device fabrication**

Thin film h-BN and few-layer graphene flakes were prepared by mechanically exfoliating high-quality bulk crystals onto a SiO$_2$/Si substrate. The layer number of graphene flakes was first identified from the optical contrast (Extended Data Fig. 1) and then confirmed directly by later transport data (Extended Data Fig. 3). The rhombohedral domains of heptalayer graphene were identified and mapped by Raman spectroscopy. The rhombohedral stacking of multilayer graphene can be easily converted into the energetically stable Bernal stacking during the van der Waals transfer process. To reduce the risk of relaxation, the rhombohedral stacking domains were isolated from the Bernal stacking domains in advance by atomic force microscopy (AFM) cutting[28,50]. To further increase the success rate, the transfer process was separated into two parts. For the bottom gate part, a bottom h-BN flake was picked up by polypropylene carbonate (PPC) film and released onto a few-layer graphite flake. The PPC was removed by acetone. The few-layer graphite/h-BN bottom part was first annealed in Ar/H$_2$ gas at 300 °C and then cleaned by AFM using contact mode. For the top part, a top h-BN was picked up by polycarbonate (PC) film and used to pick up rhombohedral heptalayer graphene. The entire stack was finally released onto the bottom gate part, forming a final heterostructure consisting of h-BN/rhombohedral graphene/h-BN/few-layer graphite. During the transfer process, the straight edge of rhombohedral graphene was intentionally aligned to the straight edge of one h-BN, forming a single alignment configuration.

After the preparation of the heterostructures, a standard microdevice fabrication procedure was employed to achieve a standard Hall bar geometry with dual-gate structures. The pattern was created by e-beam lithography. Metallic electrodes were achieved by employing two-dimensional surface contact through etching the top h-BN with reactive-ion etching and depositing Cr/Au (5 nm/60 nm). The top gate was fabricated by the evaporation of Cr/Au (5 nm/40 nm) using an additional e-beam lithography process.

**Transport measurements**

Transport measurements were carried out in a dilute refrigerator (Oxford Triton) with a base temperature of 50 mK. All electrical wires connected to the sample passed through low-temperature RC and RF filters (QDevil) fixed in the mixing chamber. A standard low-frequency lock-in technique (Zurich MFLI) combined with a current preamplifier (SR570) at a frequency of 17.77 Hz or 33.33 Hz was used to measure the longitudinal and Hall resistance of a Hall bar device. The AC excitation current was set to be 1 nA to measure the fragile states. Gate voltages were applied using a Keithley 2614B.

The bottom gate voltage $V_b$ and top gate voltage $V_t$ are converted to $n$ and $D$ following the equations: $n = (C_b \Delta V_b + C_t \Delta V_t)/e$ and $D = (C_b \Delta V_b - C_t \Delta V_t)/2\varepsilon_0$, where $C_b$ ($C_t$) are the bottom (top)-gate capacitances per unit area, $\Delta V_b = V_b - V_b^0$ ($\Delta V_t = V_t - V_t^0$) are the effective bottom (top) gate voltages, $e$ is the elementary charge, and $\varepsilon_0$ is the vacuum permittivity. $C_b$ ($C_t$) were measured from the Hall effect in normal regions and calibrated by the quantum Hall sequences (Extended Data Fig. 4).

**Symmetrized $R_{xx}$ and anti-symmetrized $R_{xy}$**

Even in standard Hallbar geometry, $R_{xx}$ and $R_{xy}$ usually mix with each other. To disengage them, we used the standard procedure to symmetrize and anti-symmetrize the measured $R_{xx}$ and $R_{xy}$. For the data shown at a fixed $B$, we used: $R_{xy}(\pm B) = [R_{xy}(B) - R_{xy}(-B)]/2$ and $R_{xx}(\pm B) = [R_{xx}(B) + R_{xx}(-B)]/2$. For magnetic hysteresis data, we employed: $R_{xy}^{\text{anti-sym}}(B, \leftarrow) = [R_{xy}(B, \leftarrow) - R_{xy}(-B, \rightarrow)]/2$ and $R_{xy}^{\text{anti-sym}}(B, \rightarrow) = [R_{xy}(B, \rightarrow) - R_{xy}(-B, \leftarrow)]/2$ ; $R_{xx}^{\text{sym}}(B, \leftarrow) = [R_{xx}(B, \leftarrow) + R_{xx}(-B, \rightarrow)]/2$ and $R_{xx}^{\text{sym}}(B, \rightarrow) = [R_{xx}(B, \rightarrow) + R_{xx}(-B, \leftarrow)]/2$, where $\leftarrow$ and $\rightarrow$ indicate the magnetic field sweep direction. Anomalous residual resistance in Fig. 3e is defined as $\Delta R_{xy}(B) = [R_{xy}(B, \rightarrow) - R_{xy}(B, \leftarrow)]/2$.

**Twist angle estimation**

The twist angle between rhombohedral graphene and h-BN can be calculated from resistance peaks at $n_1$ and $n_2$ corresponding to $\nu = 1$ and $\nu = 2$, respectively. Given the four-fold degeneracy (two for spin and two for valley) in graphene, four electrons per moiré cell are required for full filling of a moiré miniband ($\nu = 4$). The corresponding $n$ at $\nu = 4$ is $n_s = 4n_1 = 4/A$, where $A = \sqrt{3}\lambda^2/2$ is the unit-cell area of the superlattice. The moiré wavelength is $\lambda = \frac{(1+\delta)a}{\sqrt{2(1+\delta)(1-\cos\theta)+\delta^2}}$, where $a = 0.246$ nm is the lattice constant of graphene, $\delta \approx 0.0163$ and $\theta$ are the lattice mismatch and twist angle between graphene and h-BN, respectively. The twist angle calculated from this method is 0.53° for device D1.

The twist angle can also be calculated based on Brown-Zak oscillations[51]. In systems with superlattices under a magnetic field, the electronic spectra can develop into fractal spectra known as Hofstadter butterflies, resulting in a series of minimal $R_{xx}$ at $B = \phi_0/qA$, where $q$ is an integer and $\phi_0$ is magnetic flux quantum. The low resistance observed in Brown–Zak oscillation stems from the repetitive formation of magnetic Bloch states at magnetic field following the sequence of $\frac{\phi}{\phi_0} = 1/q$, in which electrons recover delocalized wave functions and propagate along open trajectories instead of cyclotron trajectories. By fitting this Brown-Zak oscillation in device D1 (Extended Data Fig. 4), we extracted the moiré wavelength of $\lambda = 13.4$ nm, corresponding to the twist angle of $\theta = 0.51°$.

**Incommensurate Wigner crystals at small fillings and $\nu = 2$**

Electron crystallization can be not only topologically nontrivial but also topologically trivial in terms of Wigner crystals. We observed a series of topologically trivial correlated insulating states at small fillings and near $\nu = 2$ (see Extended Data Figs. 7 and 8), which are incommensurate with the moiré superlattice, different from previous reports[4,7]. We find vanishing $R_{xy}$ and dispersionless $R_{xx}$ at these regions. Specifically, at $\nu = 2$, $R_{xy}$ exhibits linear behavior and vanishing hysteresis loops (see Extended Data Fig. 8) when sweeping $B$, indicating their non-ferromagnetic characteristics. Furthermore, these states are non-dispersive with magnetic fields, confirming they are topologically trivial states. Since they prefer to appear at low density close to band edge, Wigner crystal is the most likely candidate for these charge-ordered insulating states.

**Anomalous Hall crystals in device D2**

Extended Data Figs. 10 and 11 show the phase diagram of device D2 in the range of $1 \leq \nu \leq 2$, exhibiting the phase boundary resembling that of device D1. The twist angle of device D2 is 0.47°. In general, the topological states are similar to those of device D1, featuring large residual $R_{xy}$ at low $B$, hysteretic anomalous Hall effects, and dispersive Landau fan diagrams. A large anomalous resistance spans over the entire range of $1 \leq \nu \leq 2$. At a fixed $D$, the electrons crystalize into three Chern states, with $C = 3$ at $1.5 \leq \nu \leq 2$.

Nevertheless, we identify two distinct features following a meticulous comparison of device D2 and device D1. The first one is that at $\nu = 2$ topologically nontrivial states are observed in device D2. The Chern number at $\nu = 2$ is $C = 1$ extracted from the Streda slope (see Extended Data Fig. 11), again different from that of $C = 3$ at $1.5 \leq \nu \leq 2$. The second one is that the $C = 3$ states extend to a large range between $1.5 \leq \nu \leq 2$ at a fixed $D$, which agrees with AHC picture. The slightly different topological features between device D2 and device D1 indicate the twist angle of graphene/h-BN moiré superlattices may modulate the Berry curvature distribution in the second correlated band, resulting in diverse strong-interaction-driven Chern states.

**Band structure calculation based on continuum model**

The single-particle band structure was calculated by using continuum model. Specifically, the Hamiltonian of rhombohedral heptalayer graphene/h-BN moiré superlattice is

$$H_{tot} = H_7 + V_{mo}, \quad \ldots\ldots(1)$$

where $H_7$ is the effective Hamiltonian of the intrinsic rhombohedral heptalayer graphene, constructed from Slonczewski-Weiss-McClure tight-binding lattice model, and $V_{mo}$ is the effective moiré potential[52]. Mathematical details and physical parameters can be found in Ref[21].

**Hartree-Fock calculation**

The interacting band structure was calculated by further applying Hartree-Fock method to the single-particle band structure[9,10]. The Coulomb interaction Hamiltonian is given by

$$H^{int} = \frac{1}{2A} \sum_{\alpha,\alpha'} \sum_{\widetilde{k}_1,\widetilde{k}_2,\widetilde{q}} V_{\alpha\alpha'}(\widetilde{q}) c^\dagger_{\alpha,\widetilde{k}_1+\widetilde{q}} c^\dagger_{\alpha',\widetilde{k}_2-\widetilde{q}} c_{\alpha',\widetilde{k}_2} c_{\alpha,\widetilde{k}_1}$$

where $A$ is the total area of the sample, $\alpha = A_1, \ldots, B_7$ corresponds to the mixture of sublattice and layer indices, $\widetilde{k}, \widetilde{q}$ are defined in the big Brillouin zone, and

$$V_{\alpha\alpha'}(q) = \frac{e^2 \tanh(|q|d_s)}{2\varepsilon_0 \varepsilon_r |q|}$$

with the screening length $d_s = 30$ nm and dielectric constant $\varepsilon_r = 8$. The first five conduction bands were used to calculate. The details can be found in Ref[21].

**Spatial charge distribution calculation**

The Bloch wave function calculated from the Hartree-Fock method has the following form:

$$\psi_{\alpha,k;n}(r) = \sum_G u_{\alpha,G;n}(k) e^{ir \cdot (k+G)}$$

where $n$ is the band index, $k$ is defined in the moiré Brillouin zone and $G$ is the reciprocal lattice in the moiré Brillouin zone. The local density of state $\rho(r)$ is calculated by[53]

$$\rho(r) = \sum_{\alpha,n,k} |\psi_{\alpha,k;n}(r)|^2.$$

**Acknowledgements:**
This work was funded by National Natural Science Foundation of China (Grant No. 12274354, S.X.; No. 12474144, W.Z.), the Zhejiang Provincial Natural Science Foundation of China (Grant No. LR24A040003, S.X.; No. XHD23A2001, S.X.), and Westlake Education Foundation at Westlake University. We thank Chao Zhang from the Instrumentation and Service Center for Physical Sciences (ISCPS) at Westlake University for technical and facility support in data acquisition. We also thank the Instrumentation and Service Center for Molecular Sciences (ISCMS) at Westlake University for facility support. K.W. and T.T. acknowledge support from the JSPS KAKENHI (Grant Numbers 21H05233 and 23H02052) and World Premier International Research Center Initiative (WPI), MEXT, Japan.

**Author Contributions**
S.X. conceived the idea and supervised the project. H.X. and J.D. fabricated the devices with the assistance of N.L. and Q.C.. J.D. and H.X. performed the transport measurement with the assistance of N.L. and W.Zhou. J.H. and W.Zhu calculated the band structure. N.X. and W.Zhu contributed to the data analysis. K.W. and T.T. grew h-BN crystals. S.X. and J.D. wrote the paper with the input from H.X. and J.H.. All authors contributed to the discussions.

# Figures

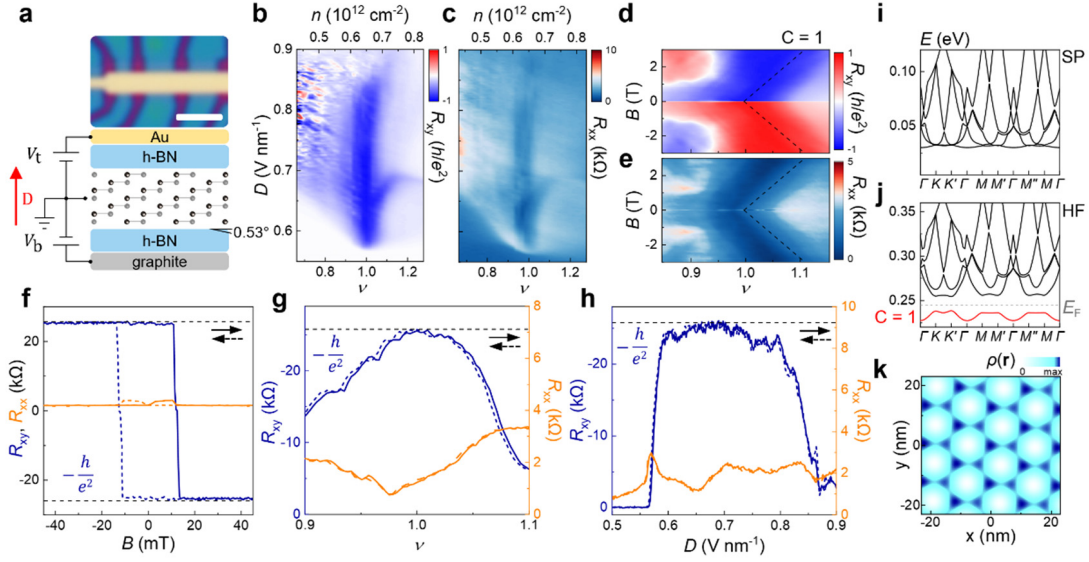

**Fig. 1 | Quantum anomalous Hall insulator at $\nu = 1$. a**, Schematic of dual-gated graphene/h-BN moiré heterostructure device. Rhombohedral heptalayer graphene was crystallographically aligned with bottom h-BN. The red arrow marks the direction of $D > 0$. The upper panel shows the final device. The scale bar is 2 μm. **b, c**, Phase diagram near $\nu = 1$ at the moiréless interface with $D > 0$ and $n > 0$. Anti-symmetrized Hall resistance $R_{xy}$ (**b**) and symmetrized longitudinal resistance $R_{xx}$ (**c**) are plotted as functions of $\nu$ and $D$ measured at $B = \pm 50$ mT. **d, e**, Landau fan of $R_{xy}$ (**d**) and $R_{xx}$ (**e**) as functions of $\nu$ and $B$ measured at fixed $D = 0.640$ V nm$^{-1}$. Dashed lines show the expected dispersions with Chern number of $C = 1$ based on the Streda formula. **f**, Magnetic hysteresis loop of QAH state at $\nu = 1$. $R_{xy}$ and $R_{xx}$ are measured by sweeping $B$ forward and backward at fixed $D = 0.640$ V nm$^{-1}$. **g**, Measurements of $R_{xy}$ and $R_{xx}$ as a function of $\nu$ swept forward and backward at fixed $D = 0.644$ V nm$^{-1}$. **h**, Measurements of $R_{xy}$ and $R_{xx}$ as a function of $D$ swept forward and backward at fixed $\nu = 1$. Data in (**g**) and (**h**) were extracted by symmetrizing $R_{xx}$ and anti-symmetrizing $R_{xy}$ taken at $B = \pm 50$ mT. All data were measured at $T = 50$ mK. **i**, The calculated single-particle band structure of a rhombohedral heptalayer graphene/h-BN moiré superlattices with a twist angle of 0.53°. An interlayer potential of 10 meV is introduced to mimic the displacement field $D$. **j**, Self-consistent Hartree-Fock band structure at the filling factor of $\nu = 1$. An isolated Chern band with $C = 1$ is observed. **k**, Spatial charge density distribution shows that an electronic crystal with a honeycomb-like lattice is formed.

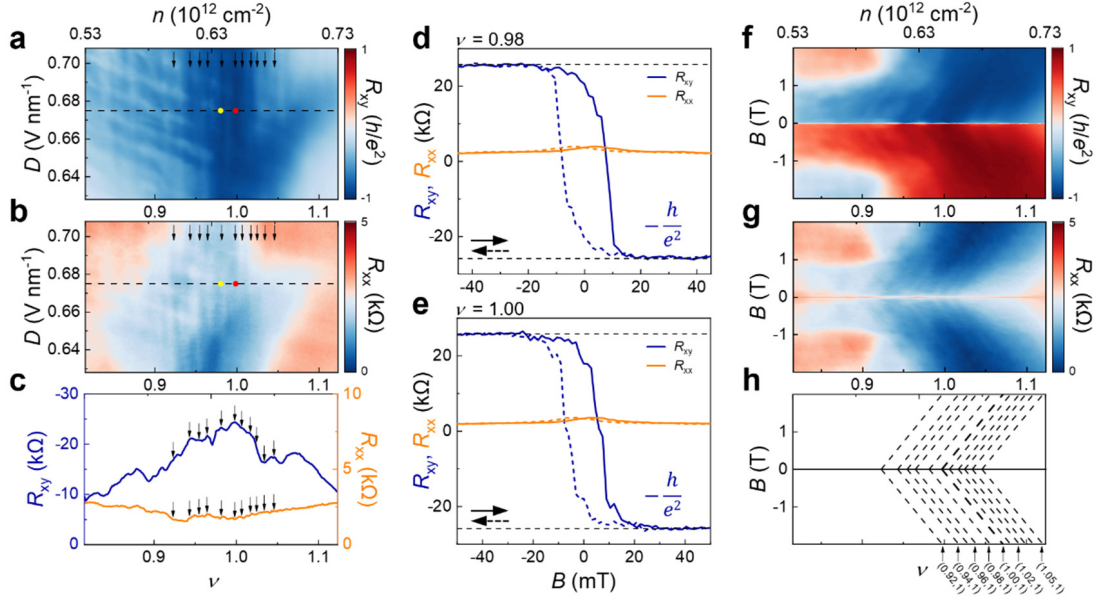

**Fig. 2 | Incommensurate Chern insulators near $\nu = 1$ at $D > 0.650$ V nm$^{-1}$. a, b,** Fine maps of anti-symmetrized $R_{xy}$ (**a**) and symmetrized $R_{xx}$ (**b**) as functions of $\nu$ and $D$ measured at large $D$ region and $B = \pm 50$ mT. The arrows mark multiple stripe features in $R_{xy}$ and $R_{xx}$. **c,** Line cut plots of $R_{xy}$ and $R_{xx}$ marked in (**a**) and (**b**) at fixed $D = 0.675$ V nm$^{-1}$. **d, e,** Magnetic hysteresis loop of $R_{xy}$ and $R_{xx}$ measured by sweeping $B$ forward and backward at two representative positions: $\nu = 0.98$ (**d**) and $\nu = 1.00$ (**e**) marked by the yellow and red dots in (**a**), respectively. The displacement fields are at fixed $D = 0.675$ V nm$^{-1}$. **f, g,** Landau fan of $R_{xy}$ (**f**) and $R_{xx}$ (**g**) as functions of $\nu$ and $B$ measured at fixed $D = 0.675$ V nm$^{-1}$. **h,** Corresponding Wannier diagram with dashed lines marking the dispersions according to the local $R_{xy}$ maxima in (**f**) and local $R_{xx}$ minima in (**g**). The indices illustrate $(\nu, C)$, where $C$ is the Chern number extracted from the Streda formula and $\nu$ is the point from which Landau fan emanates. All the Streda dispersion lines yield the same Chern number of $C = 1$.

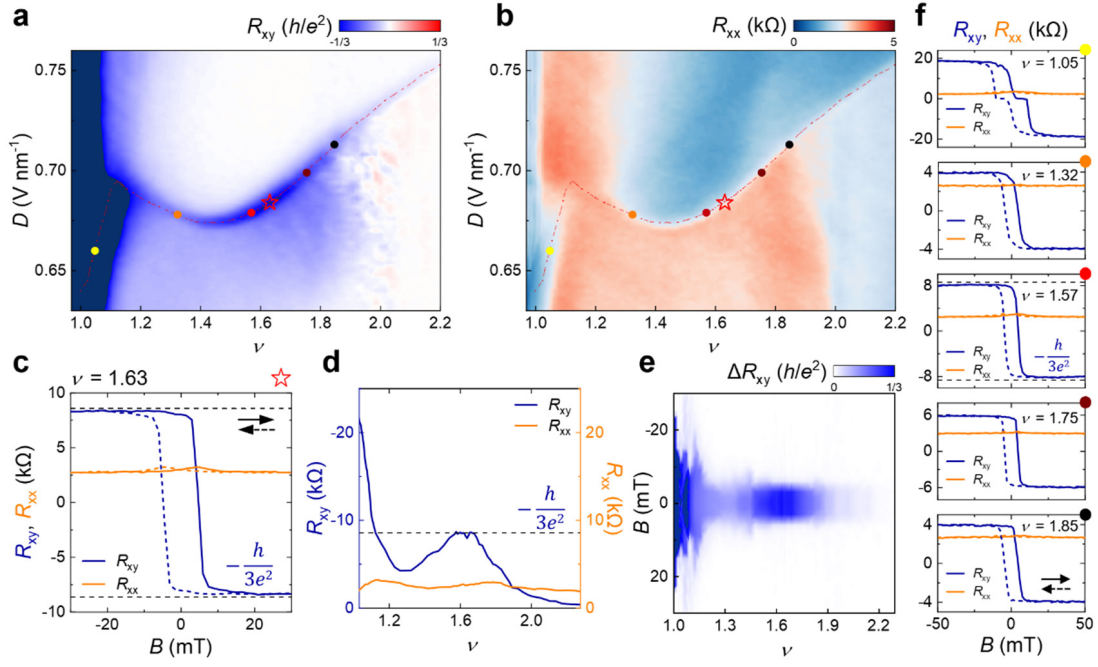

**Fig. 3 | Hysteretic anomalous Hall effects at generic fillings $1 < \nu < 2$. a**, **b**, Phase diagram at second correlated band obtained by plotting anti-symmetrized $R_{xy}$ (**a**) and symmetrized $R_{xx}$ (**b**) as functions of $\nu$ and $D$, measured at $B = \pm 50$ mT. **c**, $R_{xy}$ and $R_{xx}$ as a function of $B$ swept forward and backward at $\nu = 1.63$ and $D = 0.684$ V nm$^{-1}$. The corresponding position is marked by hollow stars in (**a**) and (**b**). $R_{xy}$ is approximately quantized at $h/3e^2$. **d**, $R_{xy}$ and $R_{xx}$ as a function of $\nu$ measured along the dashed line marked in (**a**) and (**b**). **e**, Anomalous residual Hall resistance $\Delta R_{xy}$ defined in Methods as functions of $B$ and $\nu$ measured along the dashed line marked in (**a**) and (**b**). **f**, Magnetic hysteresis loop of $R_{xy}$ and $R_{xx}$ measured by sweeping $B$ forward and backward at representative $\nu$ and $D$. For each measurement, the corresponding $D$ is chosen such that the position of $(\nu, D)$ is located at the dashed line marked in (**a**) and (**b**). The data in (**e**) are extracted from a series of magnetic hysteresis measurements as shown in (**f**).

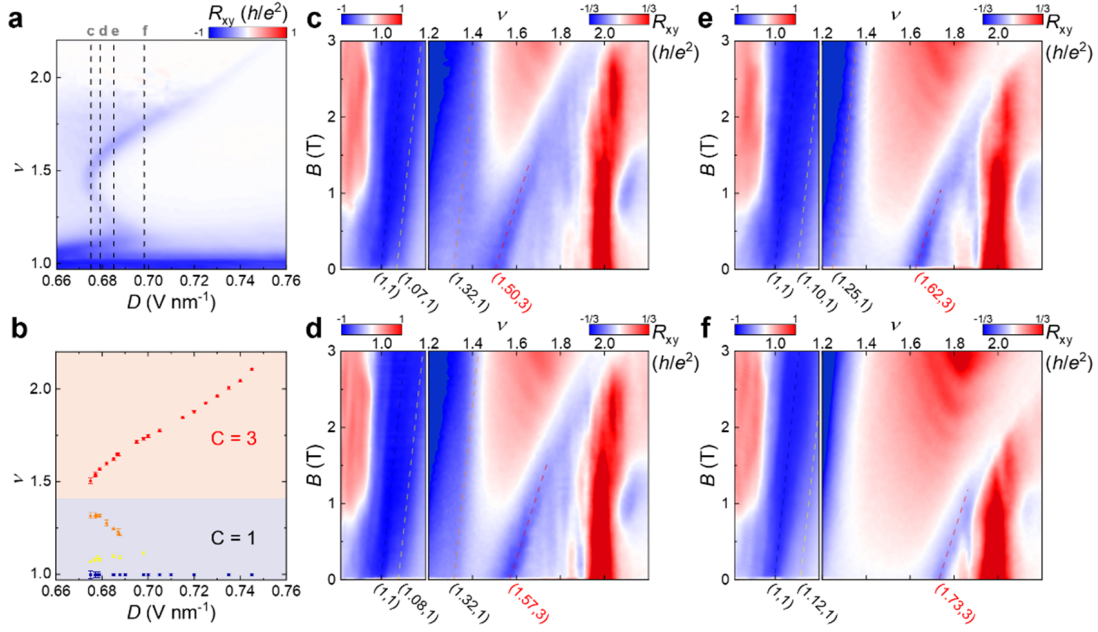

**Fig. 4 | Displacement-field tunable Chern insulators at generic fillings $1 < \nu < 2$. a**, Anti-symmetrized $R_{xy}$ as functions of $\nu$ and $D$ measured at $B = \pm 50$ mT. **b**, Four groups of Chern states plotted by $\nu$ as a function of $D$. The data with error bars were extracted from Landau fans based on the Streda formula. **c-f**, Representative Landau fan of $R_{xy}$ as functions of $\nu$ and $B$ measured at fixed $D = 0.675$ V nm$^{-1}$ (**c**), $D = 0.679$ V nm$^{-1}$ (**d**), $D = 0.685$ V nm$^{-1}$ (**e**), $D = 0.698$ V nm$^{-1}$ (**f**). The corresponding $D$ at which (**c**)-(**f**) are measured are marked by the dashed vertical lines in (**a**). The colored dashed lines in (**c**)-(**f**) illustrate the Streda dispersion lines from which the corresponding Chern numbers are extracted. The coordinates $(\nu, C)$ mark the corresponding $\nu$ at $B = 0$ from which Landan fans emanate, along with their Chern numbers $C$. The data in (**b**) are extracted from a series of Landan fan measurements as shown in (**c**)-(**f**). The distribution of the points in (**b**) is roughly consistent with the local $R_{xy}$ maxima in (**a**), suggesting that Chern states occur at entire second correlated band and are tunable by $D$.